\documentclass[11pt,
  paper=216mm:303mm,
  parskip=half,
  ngerman,
  english,
  oneside,
  listof=totoc]{scrbook}
\usepackage[ngerman, english]{babel}
\usepackage{titling}
\areaset{398pt}{596pt}
\title{Securing Network-Booting Linux Systems at the Example of bwLehrpool and bwForCluster NEMO}
\author{Simon~Moser}

\newcommand{\firstexaminer}{ Prof.~Dr.~Peter Thiemann}
\newcommand{\secondexaminer}{Prof. Dr. Christoph Scholl}
\newcommand{\advisers}{Dr.~Dirk von Suchodoletz, Michael Janczyk, Simon Rettberg, Christian Rößler, Bernd Wiebelt}
\newcommand{\mainadviser}{Dr.~Dirk von Suchodoletz}

    \usepackage{scrhack}
    \usepackage[eulergreek]{sansmath}

    \usepackage[labelfont=bf, labelsep=colon, format=hang, textfont=singlespacing]{caption}
    \usepackage{chngcntr}  
    \counterwithout{equation}{chapter}
    \counterwithout{figure}{chapter}
    \counterwithout{table}{chapter}
    \usepackage{pgf-umlsd}
    \usepackage{pgfplots}

    \graphicspath{ {./img/} }
    \newcommand{\smimage}[3][width=1\textwidth]{	%
	    \begin{figure}[h]						%
	    	\centering							%
	    	\includegraphics[#1]{#2}			%
	    	\caption{#3}						%
	    	\label{fig:#2}						%
	    \end{figure}							%
	}

    \setkomafont{chapter}{\normalfont\bfseries\huge}

    \usepackage{setspace}  
    \usepackage{xcolor}  
    \usepackage{helvet}
    
    \usepackage{array}  
    \usepackage{graphicx}  
    \usepackage{subfig}  
    \usepackage{amsmath}  
    \usepackage{amsthm}   
    \usepackage{amsfonts} 
    \usepackage{calc}  
    \usepackage[hyphens]{url}
    \usepackage[unicode=true,bookmarks=true,bookmarksnumbered=true,
                bookmarksopen=true,bookmarksopenlevel=1,breaklinks=false,
                pdfborder={0 0 0},backref=false,colorlinks=false]{hyperref}
    \usepackage{varioref}
    \usepackage{cleveref}
    \usepackage{etoolbox} 
    \usepackage[hang,flushmargin,ragged]{footmisc}
    \usepackage{csquotes}
    \usepackage[listings,breakable,skins]{tcolorbox}

    \definecolor{darkgreen}{rgb}{0.0, 0.5, 0.0}
    \definecolor{darkred}{rgb}{0.5, 0, 0.0}
    \definecolor{UniBlau}{RGB}{52, 74, 154}
    \definecolor{UniGruen}{RGB}{0, 153, 125}
    \definecolor{UniBraun}{RGB}{143, 107, 48}
    \definecolor{UniGelb}{RGB}{255, 232, 99}
    \definecolor{UniRosa}{RGB}{245, 194, 237}
    \definecolor{UniRosaStrong}{RGB}{255, 151, 171}
    \definecolor{UniSand}{RGB}{246, 241, 227}

    \lstdefinelanguage{diff}{
        morecomment=[f][\color{UniGruen}]{+},
        morecomment=[f][\color{UniRosaStrong}]{-},
    }
    \newtcblisting[auto counter, list inside=codes,crefname={lst.}{lst.s},Crefname={Listing}{Listings}]{tcbcode}[2][]{%
        skin=enhanced,
        title=Listing~\thetcbcounter:~#2,
        list entry=\thetcbcounter\qquad#2,
        listing only,
        size=small, left=6mm,
        code={\setstretch{1.125}},
        colback=UniSand!50!white,
        colbacktitle=UniBlau,
        colframe=UniBlau,
        breakable,
        lines before break=3,
        #1
    }

    \usepackage{tikz}
    \usetikzlibrary{shapes,arrows,backgrounds,fit,positioning}
    \tikzset{>=latex}

    \usepackage{enumitem}
    \usepackage{booktabs}

    \let\mySection\section\renewcommand{\section}{\suppressfloats[t]\mySection}
    \let\mySubSection\subsection\renewcommand{\subsection}{\suppressfloats[t]\mySubSection}

    \hypersetup{pdftitle={\thetitle},
                pdfauthor={\theauthor},
                pdfsubject={Master thesis at the Albert Ludwig University of Freiburg},
                pdfkeywords={bwLehrpool, bwHPC, bwForCluster NEMO, remote boot, TPM, Secure Boot, security},
                pdfpagelayout=OneColumn, pdfnewwindow=true, pdfstartview=XYZ, plainpages=false}

    \newcommand{\footurl}[1]{\footnote{See: \url{#1}}}
    
    \setlist{noitemsep}

\begin{document}
    \pagestyle{empty}
    \hypersetup{pageanchor=false}

\begin{titlepage}
\begin{center}

\vspace*{1cm}
{\Large Master's Thesis }\\[1.3cm]

{ \huge \bfseries \thetitle }

{\LARGE \theauthor} \\[2cm]

\vfill

\begin{tabular}[hc]{>{\Large}l >{\Large}l}
  Examiner: & \firstexaminer \\[0.3cm]
  Main Adviser: & \mainadviser \\[1.2cm]
\end{tabular}

\vfill

\large {
    University of Freiburg\\
    Faculty of Engineering\\
    Department of Computer Science\\
    Chair for Communication Systems\\[1cm]
	September 11, 2023\\
}
\vfill
\hspace*{-1.5cm}\vspace*{-5cm}
\begin{minipage}[b]{0.5\linewidth}
	\includegraphics[width=1\linewidth]{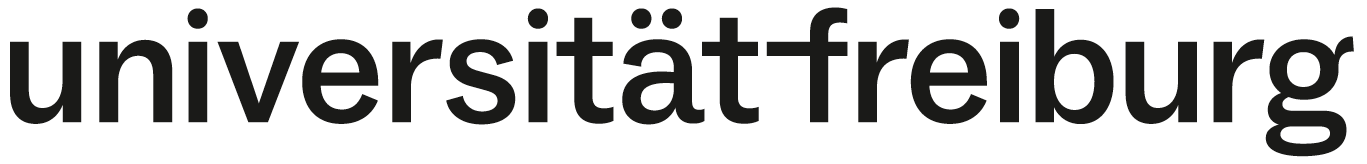}
\end{minipage}
\begin{minipage}[b]{0.5\linewidth}
	\hspace*{6cm}\includegraphics[width=0.5\linewidth]{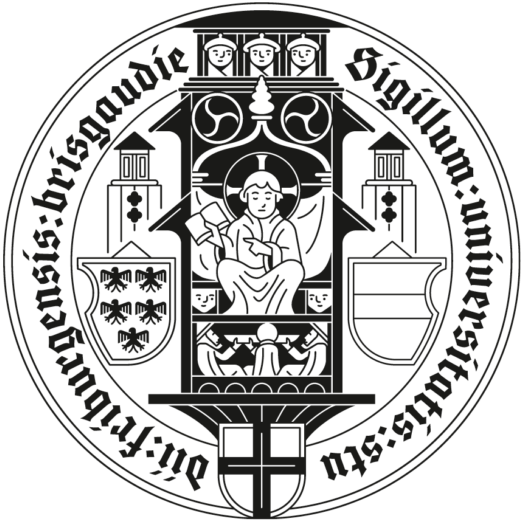}
\end{minipage}
\end{center}
\end{titlepage}

\thispagestyle{empty}
\ \vfill \ \\  
\
\textbf{Writing Period}            	\smallskip{} \\
09.\,03.\,2023 -- 11.\,09.\,2023    	\bigskip{} \\
\
\textbf{Examiner}                  	\smallskip{} \\
\firstexaminer                     	\bigskip{} \\
\textbf{Second Examiner}       		\smallskip{} \\
\secondexaminer                		\bigskip{} \\
\
\textbf{Advisers}                  	\smallskip{} \\
\advisers

    \pagestyle{headings}
    \frontmatter  
    \chapter*{Declaration}

I hereby declare, that I am the sole author and composer of my thesis and that no other sources or learning aids other than those listed have been used.
Furthermore, I declare that I have acknowledged the work of others by providing detailed references of said work. \newline
I also hereby declare that my thesis has not been prepared for another examination or assignment, either in its entirety or excerpts thereof.
\\[3\normalbaselineskip]
\begin{tabular}{p{\textwidth/2} l}
  \rule{\textwidth/3}{0.4pt}   &   \rule{\textwidth/3}{0.4pt} \\
  Place, Date                  &   Signature
\end{tabular}

    \chapter*{Abstract}
The universities of Baden-Württemberg are using stateless system remote boot for services such as computer labs and data centers.
It involves loading a host system over the network and allowing users to start various virtual machines.
The filesystem is provided over a distributed network block device (dnbd3) mounted read-only.

The process raises security concerns due to potentially untrusted networks.
The aim of this work is to establish trust within this network,
focusing on server-client identity, confidentiality and image authenticity.

Using Secure Boot and iPXE signing, the integrity can be guaranteed for the complete boot process.
The necessary effort to implement it is mainly one time at the set-up of the server,
while the changes necessary once to the client could be done over the network.
Afterwards, no significant delay was measured in the boot process for the main technologies,
while the technique of integrating the kernel with other files resulted in a small delay measured.

TPM can be used to ensure the client's identity and confidentiality.
Provisioning TPM is a bigger challenge because as a trade-off has to be made
between the inconvenience of using a secure medium and the ease of using an insecure channel once.
Additionally, in the data center use case, hardware with TPM might have higher costs,
while the additional security gained by changing from the current key storage is only little.
After the provisioning is completed, the TPM can then be used to decrypt information with a securely stored key.

    \tableofcontents
    \listoffigures
    \tcblistof[\chapter*]{codes}{List of Code listings}
    \addcontentsline{toc}{chapter}{List of Code listings}
    \hypersetup{pageanchor=true}

    \mainmatter 
    \chapter{Introduction}\label{ch:introduction}

The universities of Baden-Württemberg are actively using
and developing multiple services based on the idea of stateless system remote boot,
like bwLehrpool\footurl{https://www.bwlehrpool.de/} and bwForCluster NEMO\footurl{https://www.nemo.uni-freiburg.de/}\@.

Doing so offers various advantages:
\begin{itemize}
    \item{}The provisioning of the clients can be done fast.
    Only minimal manual provisioning is necessary.
    The images can be used for multiple or even all clients without further adjustments~\cite[section \enquote{Boot Selection Service}]{mocken2019}.
    \item{}Different images can be easily offered to the users
    (in bwLehrpool~\cite[section \enquote{Angepasste VM-Images}]{Suchodoletz2014})
    or selected automatically
    (in bwForCluster NEMO~\cite[section \enquote{Flexibility}]{mocken2019})
    \item{}No data needs to be copied to the local disk.
    The disk can be used for another operating system or for caching data~\cite[section \enquote{Systemarchitektur}]{Suchodoletz2014}.
    \item{}If data is stored on the disk for caching purposes,
    it is fully encrypted with a one-time key that is only kept in memory.
    After the client is shut down, no disk access is possible anymore~\cite[section 3.2.3]{Suchodoletz2021}.
\end{itemize}

The rough sequence of this network boot at the example of bwLehrpool is as follows:
initially, the UEFI is loading the iPXE environment
which then loads a kernel and initramfs over HTTP with specific configuration for this client.
When this minimal image is loaded, the user can log in and has a selection of various virtual machines
to run depending on his own groups and on the classification of the used client.
Further data is loaded using a distributed network block device (dnbd3) developed specifically for this purpose.
This procedure is intentionally open in order to be able to integrate any system very quickly.

In this process, various secret information is transmitted between server and client
over a network that potentially cannot be trusted.
Examples of sensitive data that should be transmitted securely are configuration files,
SSH host keys or password hashes as in \texttt{/etc/shadow}, depending on the server configuration.
Therefore, the process should be changed to achieve a trust relationship
between the server and the client within a zero-trust network.
This task can be seen as two different parts,
ensuring the identity of the server and the authenticity of the transmitted images on the one hand (what is booted)
and ensuring the identity of the client and the confidentiality of potential secrets, on the other hand (who has access).

For this, however, the special requirements of the environment must be considered.
In general, the advantages listed at the beginning of this section should not be lost.
Changes to the process must ensure that a larger number of systems can continue to be managed efficiently.

\section{Related work}\label{sec:related-work}
This possible improvement was already described in a previous paper~\cite{Suchodoletz2021}.
Von Suchodoletz et al. describe the security gains of using stateless system remote boot there.
In the evaluation, the need for signing and encryption is explained.
Additionally, they also name the missing possibility for computer forensics as a motivation to create a secure channel.
With such a channel, the one-time disk encryption key could be transferred to a server to allow a forensic analysis.

    \chapter{Background}\label{ch:background}
This chapter includes, on the one hand, an overview of the architecture of the Stateless System Remote Boot,
which is the main content of this thesis, and, on the other hand, two important use cases of this system.
Furthermore, the last section introduces and explains technologies and terminology that are important for this work.

\section{Stateless System Remote Boot architecture}\label{sec:remote-boot-architecture}
As initially described by Schmelzer et al.~\cite{Schmelzer2011}, Stateless System Remote Boot
(which is called Universal Remote Boot there) is a process that is designed to allow booting systems over the network.
Compared to the traditional setup using only PXE, DHCP, and TFTP,
it allows a higher flexibility with an easier configuration.

The following Figure shows the overview of the system architecture that is given in the paper:

\smimage{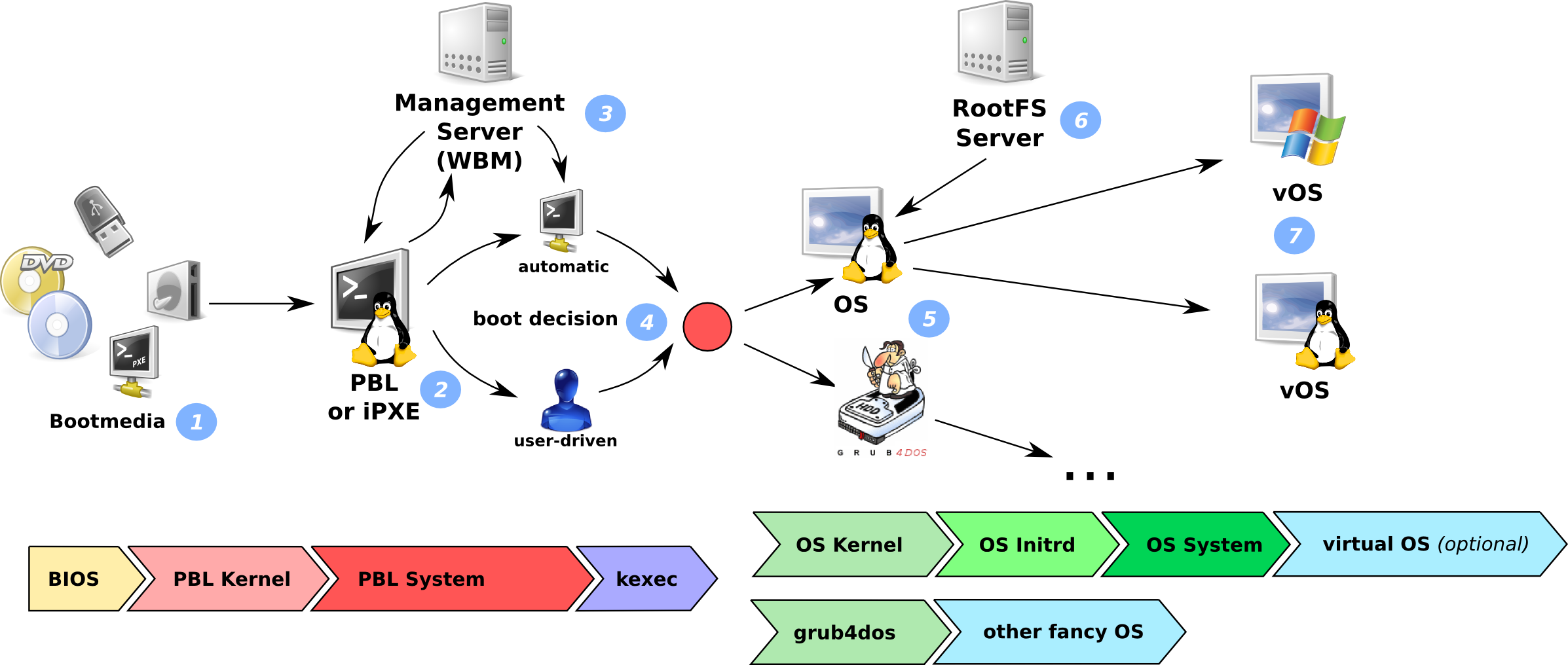}{Stateless System Remote Boot architecture~\cite{Schmelzer2011}}

\begin{itemize}
    \item In \textbf{step~1}, any boot media can be used to load the next step, the iPXE\@.
    This includes the Preboot eXecution Environment (PXE) over the network, but is not limited to it.
    \item \textbf{Step~2} is the open source implementation iPXE, which offers more options over PXE, for example scripting.
    Historically, at the time the architecture was created, a PreBootLinux (PBL) was used for this purpose instead.
    \item From the iPXE, in \textbf{step~3}, an iPXE script is loaded from a management server
    that can be created for each client and that sets up possible boot options.
    \item \textbf{Step~4} is a boot decision that can be made either by the user or automatically.
    \item On the one hand, the system can be booted into the operating system installed on the disc.
    On the other hand, as \textbf{step~5}, it can be booted into a basic Linux system,
    which will be called \enquote{host system} in the rest of this work.
    \item Within this Linux, necessary files could be retrieved from a root file system server as \textbf{step~6}.
    In the discussed use cases, this is done using distributed network block devices over the dnbd3 protocol~\cite{Rettberg2019}.
    \item In \textbf{step~7}, finally, the host system can be used to launch various virtual machines.
\end{itemize}

As already described in the introduction, the universities of Baden-Württemberg,
especially of Freiburg, are offering services based on this architecture.
The following section discusses the two use cases of the architecture that will be content of this thesis.
\newpage

\section{Stateless System Remote Boot use cases}\label{sec:remote-boot-use-cases}
\subsection{Computer labs and classrooms}\label{subsec:computer-labs}

The first use case discussed in this thesis is bwLehrpool~\cite{Suchodoletz2014,Trahasch2015,Bauer2019,Suchodoletz2022}.
It offers a system to provide computers with different operating system images
while having a minimal necessary manual configuration for enrollment and setup.
In this system, a satellite server takes the role of management server and root file system server.
It can control clients in a classroom, computer lab or even for digital exams~\cite{Ritter2016, Suchodoletz2020}.

This use case represents an open environment that is difficult to control
due to possible physical access and hard-to-control virtual machines.
The used hardware is also very heterogeneous, hence the used software must have a bigger flexibility.

In bwLehrpool, no encryption for either configuration files or disk images is used yet.

\subsection{Data centers}\label{subsec:data-centers}

The second use case for Stateless System Remote Boot discussed in this work are infrastructure services run by the
eScience department of the computer center in Freiburg, namely bwForCluster NEMO and bwCloud~\cite{Bauer2018}.

bwForCluster NEMO is a service for high-performance computing (HPC) for the researchers and students of the universities in Baden-Württemberg.
The NEMO cluster is available for research in the areas of neuroscience, elementary particle physics, microsystems engineering and material science~\cite{janczyk2018,janczyk2019}.

bwCloud is offering Infrastructure as a Service (IaaS) for science and education in four university data centers~\cite{dulov2015}.
In Freiburg, the operational concept is based on stateless system remote boot\footurl{https://www.bw-cloud.org/de/betriebskonzept/konzept_freiburg}.

Since the physical hardware is run in data centers, physical access is not an issue here.
The hardware used in data centers is more homogenous than in bwLehrpool,
but the used server hardware often lacks a Trusted Platform Module (TPM).

In the data center environment, encryption is already in use for configuration files.
The configuration is downloaded via HTTP as an encrypted archive and then decrypted using GPG2.
Due to the lack of a dedicated secure storage, the decryption key is stored
within a Data Center Manageability Interface (DCMI) asset tag~\cite[section 6.4.1]{intelcorporation2011}.
From within the system, this tag can be requested using the Intelligent Platform Management Interface (IPMI).
Access to this interface requires root privileges on the host system.

\section{Security terminology and technologies}\label{sec:bg-currentsec}
The following section comprises an explanation of the information security related terminology used in this thesis.
Also, two security technologies are described that aid the goal of zero-trust network boot.

\subsection{Information security goals}\label{subsec:information-security-goals}
This work focuses mainly on the three information security goals of confidentiality, integrity and authenticity, which are described below.
The goal of availability, which is considered one of the three main information security goals together with confidentiality and integrity~\cite[section 3.28]{iso2018},
will not be discussed in this work as the used security technologies cannot give any guarantees for availability.

\subsubsection*{Confidentiality}
The standard ISO/IEC 27000:2018 defines confidentiality as the
\enquote{property that information is not made available or disclosed to unauthorized individuals, entities, or processes}~\cite[section 3.10]{iso2018}.
This goal is usually achieved through encryption, as only someone in possession of the key can decrypt the information.
For this, the key must only be accessible to those who are authorized to access the information.

Confidentiality is only necessary if the transmitted information contains secrets that need to be protected.

\subsubsection*{Integrity}
In the ISO/IEC standard, integrity is defined as the \enquote{property of accuracy and completeness}~\cite[section 3.36]{iso2018}.
This signifies it must be guaranteed that transmitted information arrives at the receiver exactly as the sender intended it to.
Parts of the information must neither be modified, omitted nor added.
Commonly, this goal is achieved by signing a hash of the information, as any change in the information also changes the hash and therefore renders the signature invalid.

\subsubsection*{Authenticity}
Authenticity is defined by the standard as the \enquote{property that an entity is what it claims to be}~\cite[section 3.3]{iso2018}.
This means that the identity of the entity must be proven.
This can be done, among other things,
by proving knowledge, for example via a shared secret,
or possession, for example of a private key by signing information.

Whenever the goal of authenticity is required in this work, integrity is also included,
because proving the authenticity of a sender is only valuable if it is also proven that the transmitted information was not tampered with.

\subsection{Trusted computing terminology}\label{subsec:trusted-computing-terminology}
\subsubsection*{Root of Trust (RoT)}
According to the Trusted Computing Group, a Root of Trust (RoT) is a \enquote{component that must always behave
in the expected manner because its misbehavior cannot be detected}~\cite[section 4.76]{tcg2019}.
An RoT needs to be trusted without any proof since it is not possible to verify it directly.
Trust in the RoT can therefore only be strengthened by circumstances such as standardization or trust in the manufacturer.
Based on a trusted RoT, which is also called trust anchor,
a chain of trust can be established by validating each link by the previous link.
One chain of trust is formed by digital certificates, which use certificate authorities (CAs) as the RoT\@.
The trust relationship provided by an RoT is always specified to a certain functionality,
some of which are explained in the following paragraphs.

\subsubsection*{Root of Trust for Storage (RTS)}
A Root of Trust for Storage (RTS)~\cite[section 9.4.2]{tcg2019} guarantees that all sensitive information stored based on it is kept secret.
One example for an RTS could be a (symmetric or asymmetric) encryption key
which can then encrypt other encryption keys (chain links) or directly encrypt any sensitive information.

\subsubsection*{Root of Trust for Reporting (RTR)}
A Root of Trust for Reporting (RTR)~\cite[section 9.4.3]{tcg2019} guarantees that reported information originates from a trusted source.
This can be achieved, for example, by signing reported information by the RTR\@.
A chain of trust can be formed by signing a public key or a certificate of a subordinate entity
by the RTR or the preceding chain link, respectively.
An RTR is not to be confused with a Root of Trust for Measurement (RTM),
as the RTR does not perform any measurements itself and is thus dependent on trust in the RTM\@.

\subsubsection*{Root of Trust for Measurement (RTM)}
A Root of Trust for Measurement (RTM)~\cite[section 9.4.1]{tcg2019} guarantees confidence that associated measurements are correct.
There are two possibilities when measuring a boot process:
\begin{itemize}
    \item A \textbf{Static Root of Trust for Measurement (SRTM)} is found in the firmware (BIOS or UEFI)
    and measures each part of the boot chain before it is executed.
    \item A \textbf{Dynamic Root of Trust for Measurement (DRTM)} is located in the processor
    and provides measurements of code as it is executed.
\end{itemize}

\subsection{Trusted Platform Modules (TPMs)}\label{subsec:trusted-platform-modules}
Trusted Platform Modules (TPMs)~\cite{tcg2019} are hardware modules in modern computers
that are meant to provide trusted computing features.
They can be integrated in the system on a separate chip (discrete TPM) or in the processor (firmware TPM).
TPM are widely spread in end user computers
(inter alia because Windows 11 requires an active TPM) but not as common in server hardware.

For trusted computing, TPMs offer two main features: Root of Trust for Storage (RTS) and Root of Trust for Reporting (RTR).

First of all, the RTS ensures that stored secrets are kept secret by denying access from outside the TPM\@.
Since the TPM only has limited storage capabilities, the RTS can protect secrets stored outside the module
by encrypting them with a key stored within the module and therefore creating a chain of trust,
which is called binding.

Secondly, the RTR provides information about whether the system has a certain state.
It has to be noted that the TPM as an RTR cannot perform measurements itself
and therefore relies on a Root of Trust for Measurement (RTM) in the processor (DRTM) and UEFI (SRTM).

The measurements done by the RTM are sent to the \textbf{Platform Configuration Registers (PCRs)} which are located in the TPM\@.
These registers only allow changes by hashing a new measurement together with the present hash.
This creates a chain of hash values where any change to a link will result in a change to the resulting hash value.

After boot, with so-called remote attestation, the TPM can issue reports about the PCR state signed by its key for reporting in his function as RTR\@.
Furthermore, the TPM can be configured to only allow access to a secret whose confidentiality is guaranteed as RTS
when a certain PCR state is given — this is called sealing.

In \Vref{fig:tpm}, an overview of the TPM architecture is shown
which illustrates the relationships between the individual RoTs located within the TPM and in other hardware modules.
It displays the TPM functionalities of \enquote{remote attestation} (provide trusted information about the system's state) in its role as RTR,
\enquote{binding} (en- and decrypt data using a key stored in the TPM) in its role as RTS and
\enquote{sealing} (allow use of a key only when the system has a certain state) which combines both roles.

\tikzstyle{line} = [draw,-latex']
\tikzstyle{box} = [rectangle,draw,text centered]
\tikzstyle{membox} = [rectangle,draw,text centered,box,minimum width=6em,minimum height=3em,align=center,node distance=7em]
\tikzstyle{circle} = [ellipse,text centered]
\tikzstyle{venncircle} = [ellipse,inner sep=0mm,fill opacity=0.5,fill=#1]
\tikzstyle{functext} = [minimum width=6em,minimum height=3em,align=center]
\begin{figure}[h!]
\begin{center}
\begin{tikzpicture}[node distance = 2cm, auto]
        \node [] (uefi-label) {UEFI};
        \node [circle,fill=UniGelb,text opacity=1,fill opacity=0.5,below of=uefi-label,node distance=0.7cm] (srtm) {\textbf{SRTM}};
        \node [box,thick,rounded corners,fit=(uefi-label) (srtm)] (uefi) {};
        \node [left of=uefi-label,node distance=7em] (cpu-label) {CPU};
        \node [circle,fill=UniRosa,fill opacity=0.5,text opacity=1,below of=cpu-label,node distance=0.7cm] (drtm) {\textbf{DRTM}};
        \node [box,thick,rounded corners,fit=(cpu-label) (drtm)] (cpu) {};
    \node [fit=(uefi-label) (cpu-label)] (rtms) {};

    \node [below of=rtms] (tpm-label) {TPM};
    \coordinate [right of=tpm-label,node distance=7em] (hidden);

    \node [below of=hidden,node distance=0.7cm] (tpm-memory-label) {TPM Memory};
    \node [membox,below of=tpm-memory-label,node distance=1cm] (enc-key) {\textit{Encryption}\\\textit{keys}};
    \node [membox,left of=enc-key] (pcrs) {\textit{PCRs}};
    \node [membox,left of=pcrs] (sig-key) {\textit{Signing}\\\textit{keys}};

    \node [functext,below of=sig-key] (remote-attestation) {remote\\attestation};
    \node [functext,below of=pcrs,node distance=3cm,minimum width=3em] (sealing) {sealing\\ };
    \node [functext,below of=enc-key] (binding) {binding\\ };

    \node [below of=remote-attestation,node distance=2cm] (rtr-label) {\textbf{RTR}};
    \node [below of=binding,node distance=2cm] (rts-label) {\textbf{RTS}};

    \begin{pgfonlayer}{background}
        \node [rectangle,fill=UniSand,fit=(tpm-memory-label) (pcrs) (sig-key) (enc-key),node distance=0.7cm] (tpm-memory) {};
        \node [venncircle=UniBlau,fit=(remote-attestation) (sealing) (rtr-label)] (rtr) {};
        \node [venncircle=UniGruen,fit=(sealing) (binding) (rts-label)] (rts) {};
        \node [box,thick,rounded corners,fit=(tpm-label) (tpm-memory) (rts) (rtr)] (tpm) {};
    \end{pgfonlayer}

    \path [line] (srtm) -- (pcrs);
    \path [line] (drtm) -- (pcrs);
    \path [line] (pcrs) -- (remote-attestation);
    \path [line] (sig-key) -- (remote-attestation);
    \path [line] (enc-key) -- (binding);
    \path [line] (pcrs) -- (sealing);
    \path [line] (enc-key) -- (sealing);
\end{tikzpicture}
\caption{Overview of the TPM architecture}
\label{fig:tpm}
\end{center}
\end{figure}
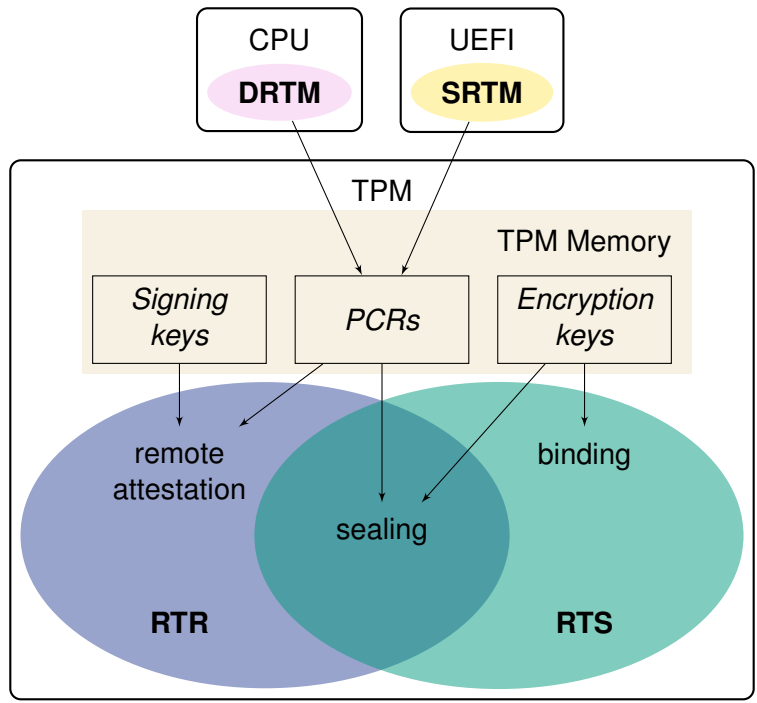
\newpage

\subsection{Secure Boot}\label{subsec:secure-boot}
Secure Boot is a part of the UEFI specification~\cite[section 32]{uefi2022}
that is supposed to guarantee the authenticity and integrity of important software.
Critical parts from the bootloader to kernel modules shall only be executed if they have been authorized.
The requirement for cryptographic signatures prevents untrusted software parts from being executed.
The secure boot process is based on a chain of trust
which means that every critical software part that will be loaded has to be signed.
These signatures can either be done using a Microsoft key as the RoT or by enrolling custom keys.
Secure boot can support ensuring the security of the keys stored in a Trusted Platform Module (TPM)
by denying unauthorized software to boot that could otherwise malicously access the TPM.

Secure boot uses four different key types:
\begin{itemize}
    \item The \textbf{Platform Key (PK)} is the Root of Trust for the system.
    It is usually issued by the manufacturer but can be replaced by a custom key.
    The PK is used to only sign the Key Exchange Key.
    \item The \textbf{Key Exchange Key (KEK)} is used to sign all further keys
    to be able to dynamically change all further keys without enrolling a completely new PK into the UEFI\@.
    \item A \textbf{Signature Database (db)} contains keys that are used to sign the verified binaries as well as hashes of such.
    For this purpose, at least one signature database key (DB) is created.
    \item A \textbf{Forbidden Signatures Database (dbx)} contains keys and hashes of binaries that are forbidden to load.
\end{itemize}

\subsection{Systemd-stub}\label{subsec:systemd-stub}
An EFI stub enables a Linux kernel to be booted by the UEFI directly, without the help of an additional bootloader like grub.
Most modern kernels contain such an EFI stub, for Debian it is in use since release \enquote{Wheezy} (2016)\footurl{https://wiki.debian.org/EFIStub}.

It is possible to combine a stub together with the kernel, an initramfs and other resources like the kernel parameters into a single binary image,
also called \textbf{Unified Kernel Image} (UKI)\footurl{https://uapi-group.org/specifications/specs/unified_kernel_image/}.
This UKI can then be signed with Secure Boot, so the full binary's integrity and authenticity can be protected by Secure Boot.
The reference implementation for a stub that can be used to create a UKI is the systemd-stub\footurl{https://manpages.debian.org/bookworm/systemd/systemd-stub.7.en.html}.

    \chapter{Problem description and requirements}\label{ch:problem}
In this chapter, the first section describes a threat scenario that shows the possibilities of an attacker.
These are further differentiated depending on the use case.
In the following section,
two vulnerabilities of Stateless System Remote Boot are explained that are based on these possibilities.
Finally, the requirements for a solution to the problems described are listed,
differentiated according to different perspectives on the system.

\section{Threat scenario}\label{sec:threat-scenario}
The threat scenario used to analyze the following attack vectors includes an attacker
that has the possibility to access the network and to get in a man-in-the-middle position,
where he can read and manipulate the traffic between server and client.
Additionally, the scenario includes all actions a typical user of the services can do on the client system.
For the general threat scenario, unsupervised physical access is also included.
However, depending on the use case, the actual threat scenario differs:

\subsection{Computer lab use case}\label{subsec:threat-computer-lab-use-case}
In the computer labs, the services are provided to the public.
Lab rooms might be open all around the clock to students and to anyone during the normal business hours.
Therefore, any attacker potentially has unsupervised physical access to the devices as well as the network.
Network access control (NAC) enforced by software is either not effective
(for example access control based on the MAC address) or requires information stored on the device
(for example the certificate-based standard IEEE 802.1X\footurl{https://1.ieee802.org/security/802-1x/}).
A working possibility for NAC would be to filter malicious packets
from the network traffic of the virtual machines in the host system.
However, this option would only limit the risk from compromised virtual machines as explained below.
Currently, this option is not used.

The lecturers can create virtual machine images which are therefore hard to manage
in a sense of monitoring and fixing security vulnerabilities.
This facilitates an attacker to gain root privileges on the virtual machines.

Altogether, an attacker has physical access to the clients and can impersonate servers and clients without big effort.

\subsection{Data center use case}\label{subsec:threat-data-center-use-case}
In the data center, physical access control measures are used, for example,
to prevent unauthorized persons from entering rooms or accessing locked server cabinets or switches.
Therefore, any attacks based on direct physical access can be disregarded here.
Access to the nodes is only provided remotely over a small number of login nodes.

Also, the images run on the nodes are easier to manage than in the bwLehrpool environment.
Inside the virtual machine, a user has root privileges, but cannot change the configured VLAN tag as it is enforced on the host.
Therefore, vulnerabilities of the virtual machines are not an issue in this use case.

Access to the different network segments is controlled by configured VLANs.
However, by exploiting the hypervisor,
an attacker might still be able to freely access the network by VLAN hopping~\cite[Section 5.1]{Bauer2018}.
By doing this, it might be possible to impersonate clients and servers.

The IPMI used to access the decryption key currently stored in an asset tag, however,
might contain vulnerabilities that allow an unauthorized attacker to access this tag,
as the asset tag is not designed for secure storage and the management software is rarely updated.

\section{Attack vectors}\label{sec:pb-attacks}

The current network boot infrastructures have two main security problems that have to be approached each separately.
Those problems will be described in the following sections.

\subsection{Malicious server}\label{subsec:pb-attacks-server}

One the one side, missing server and message authentication are leading to the fact that an attacker can act as a server or modify the response from the server.

Server authentication in this case means that the client can verify the identity of the sending server, for example by a signature or a shared secret.

Message authentication here means that the integrity of messages originating from a legitimate sender is checked to prevent individual parts of the transmitted messages from being exchanged, omitted or added by an attacker.

A manipulated image such as the initramfs for the host system can be used for multiple attacks.
First of all, an attacker could gain unprivileged access to said host system, which could be used for further steps.
Then he could extract any information stored in the client's hardware, for example,
such as currently the encryption key for the config file which is stored in the DCMI in the data center use case.
Additionally, he could gain access to any information a legitimate user using the attacked computer is inputting,
by logging the keyboard input, for example.

This kind of attacks is possible at any point of the network boot.
Possible attacks for the different transport protocols are listed in the following.
A compromised step includes the possibility to compromise all following steps since every step relies on the outcome of the previous step.

\begin{itemize}
	\item \textbf{DHCP}: send user to a malicious TFTP server for the iPXE boot image
	\item \textbf{TFTP}: send a malicious iPXE configuration or image
	\item \textbf{HTTP}: send a malicious personalized iPXE config, exchange the kernel, the initramfs, kernel parameters etc.
	\item \textbf{DNBD}: exchange certain blocks e.g.\ to allow privilege escalation or exfiltrate sensitive information
\end{itemize}

\subsection{Malicious client or passive eavesdropper}\label{subsec:pb-attacks-client}

On the other side, without client authentication and effective encryption,
an attacker can gain access to the transmitted data by passively reading all information in clear text
or by simulating a client and reading information from the received images.

Client authentication means here that the server verifies the identity of the client that receives the information,
for example by previous authentication or encrypting with a shared secret.

Effective encryption signifies that the transmitted data is encrypted in a way that an attacker who is able to read
all communication but has no further private knowledge is not able to decrypt the data and that the key
used for the encryption is not retrievable for an attacker that uses the means described in the threat scenario.

Currently, in the data center use case, a configuration file transmitted over HTTP is encrypted by using a shared static key.
However, the key is stored in the client's hardware tags which means that an attacker is able to extract this key if
he gains privileged access to the host system, by the attacks described in the previous section, for example.

The information an attacker can extract from the transmitted information over the dnbd3 protocol includes,
for example, password hashes or personal information which is protected by the GDPR\@.

The preceding transport protocols like DHCP, TFTP and HTTP mainly leak non-critical information
as long as no encryption key for further stages is transmitted,
except for the encrypted configuration in the data center use case.

\section{Requirements specification}\label{sec:pb-requirements}
All approaches to solving the previously described attacks must fulfil several requirements derived from the nature of the projects.
The requirements are also based on the advantages listed in \Vref{ch:introduction} and differ depending on the perspective from which they are viewed.
The perspectives of the users, the administrators and, in the case of the computer lab use case, the lecturers must be differentiated.

In particular, the requirements from the view of an administrator are as following:
\begin{itemize}
	\item Previously mentioned vulnerabilities should be fixed,
	especially to prevent unauthorized access to a large number of client or server systems.
	\item The process of deploying a new client should be with
	as little manual effort as possible to reduce IT administrative overhead.
	Optimally, a client just has to be connected to the right network,
	added to a client list on a central server and can directly run.
	\item After deployment, a client should not require further modifications regularly;
	so a larger number of systems can continue to be managed efficiently.
	\item Any solution must not rely on (persistently) using the local disk.
\end{itemize}

Furthermore, a user has the following requirements for the system which may partly differ from those of an administrator:
\begin{itemize}
	\item Especially all vulnerabilities affecting the user's personal data must be closed.
	\item Sensitive information as user data or key material must not be stored without protection.
	\item Furthermore, start times should not increase significantly.
	The dnbd3 protocol is designed for fast block access,
	and this goal should not be counteracted by overly big encryption overhead.
\end{itemize}

Last but not least, lecturers have additional interests:
\begin{itemize}
	\item It should be possible to create virtual machines for courses and make them available to the students
	without a high effort.
	\item The possibilities offered by different images must not be limited
	beyond the already existing boundaries such as storage space.
\end{itemize}

    \chapter{Approaches}\label{ch:approaches}
For the following chapter, the boot process in \Vref{fig:remote-boot} is divided into several sections,
each comprising one or more of the steps shown in the architecture diagram.
For each section, different approaches are described, where applicable, including their advantages and disadvantages.
This also includes the effort required to implement the solution approach.

\section{Boot to iPXE}\label{sec:boot-to-ipxe}
This part includes steps 1 and 2 of the diagram in \Vref{fig:remote-boot}.
During these steps, no confidential data is transmitted and therefore only the security goals
authenticity of the server and integrity of the data are set.

\subsection{PXE boot (original)}\label{subsec:pxe-boot}
In the original description, the client receives the initial configuration for PXE boot over the DHCP protocol~\cite[sections 9.4 and 9.5]{rfc2132}.
Using the options \texttt{Server-Name} (number 66) and \texttt{Bootfile-Name} (number 67), the client is instructed to boot a file from a given location.
The boot file is loaded by the UEFI or BIOS via the TFTP protocol.~\cite{rfc1350}

This procedure does not use any mechanism to ensure the identity of any side or the integrity or confidentiality of the data.

\subsection{UEFI HTTP boot}\label{subsec:uefi-https-boot}
With UEFI HTTP~\cite[section 24.7]{uefi2022} boot it is possible to either continue using DHCP
to request the configuration (using option 67 \texttt{Bootfile-Name} and option 60 \texttt{Vendor class identifier})
or to store the boot file location in the UEFI's storage.
In that case, the UEFI setup needs to be protected with a password in the computer lab use case to prevent manipulation of said address.
Depending on this choice, the DHCP step (step 1) of the original sequence could be completely omitted.
Step 2 will be replaced by a download of the boot file using the HTTP protocol.
This request can optionally be secured using TLS (so HTTPS).
Furthermore, it is also possible to add a custom root certificate for TLS in case that it is not possible to use a publicly valid certificate.
This feature is available in UEFI 2.5 and newer.

In case modifications of the UEFI have to be done, this can be done by manual configuration or by using an EFI startup script.
However, a change in the boot server address or the boot file location would then require updating the UEFI on every client.

With all optional measures in place, this approach could guarantee the identity of the server as well as the integrity and confidentiality of the data.

\subsection{Secure Boot}\label{subsec:pxe-secure-boot}
Both PXE boot and UEFI HTTP boot can be secured using Secure Boot as previously described in \Vref{subsec:secure-boot}.
To use it, all binaries started by the UEFI — as the iPXE binary — must be signed by
either a Microsoft signing key or by custom key pairs enrolled into the UEFI\@.

Secure boot does not pose a guarantee of the confidentiality of the transmitted data,
only for the identity of the server and the integrity of the data.

\subsection{Custom NIC firmware}\label{subsec:custom-nic-firmware}
Alternatively, it is possible to flash custom firmware onto the network interface card (NIC)\footurl{https://ipxe.org/howto/romburning}.
The firmware has to be modified specifically for every device to use iPXE instead of PXE for network booting.
Then it can be flashed onto the NIC which requires physical access to the device.
Every device and vendor might require different modifications and a different flashing process with potentially different tools.
The administrative effort resulting from the entirety of the requirements makes this option unsuitable for large scale use.

By doing this, however, the need for steps 1 and 2 can be completely removed.
Therefore, no security goals need to be fulfilled as no network connection is used.

\section{Host system boot from iPXE}\label{sec:host-system-boot}
This section comprises steps 3 and 4 of the architecture diagram.
The main security goals here are the same as in the previous section,
namely the server's authenticity and the integrity of the data.
Sensitive configuration files that would require confidentiality too will be discussed in the next section.

\subsection{HTTP (original)}\label{subsec:ipxe-http)}
Here, a configuration script personalized for the actual client is loaded that then downloads and executes the chosen kernel and initramfs.
All downloads use an unsecured HTTP connection.

\subsection{HTTPS}\label{subsec:ipxe-https}
Alternatively, both the configuration and kernel and initramfs can be loaded over an HTTPS connection initiated by iPXE\@.
A custom root certificate can be used if no public domain name is available.

However, the TLS configurations supported by iPXE\footurl{https://ipxe.org/crypto} are obsolete or even insecure.
Of the TLS versions supported, only TLSv1.2 is still considered reasonably secure\footurl{https://datatracker.ietf.org/doc/html/rfc8996}.
All supported key exchange methods are obsolete and are about to become deprecated\footurl{https://datatracker.ietf.org/doc/draft-ietf-tls-deprecate-obsolete-kex/}.
Additionally, the offered CBC mode is prone to padding oracle attacks\footurl{https://blog.cloudflare.com/padding-oracles-and-the-decline-of-cbc-mode-ciphersuites/}
and the SHA-1 algorithm is long broken\footurl{https://www.nist.gov/news-events/news/2022/12/nist-retires-sha-1-cryptographic-algorithm}.
Enabling obsolete or insecure protocols and ciphers would weaken the security gained by using HTTPS\@.
Additionally, the supported ciphers are not enabled in modern web server by default anymore,
therefore additional administrative effort on the server side would be necessary to make the use possible.

If HTTPS was correctly used, it could guarantee the authenticity of the server as well as confidentiality and integrity of the transmitted data.

\subsection{Secure Boot}\label{subsec:ipxe-secure-boot}
Secure boot, as described in \Vref{subsec:pxe-secure-boot} can also be used for this step.
In the easiest configuration, only the kernel needs to be signed.
This would guarantee the authenticity of the sending server as well as the integrity of the kernel.

However, Secure Boot will not protect the customized iPXE script for every client.
This will be discussed in \Vref{subsec:ipxe-scripts}.
To ensure the integrity of the initramfs which is loaded in the iPXE script additionally to the kernel, there are multiple possibilities:

\subsubsection{Initrd encrypted with TPM key}
First of all, the initramfs could be encrypted by the server using a key stored in the clients TPM\@.
The difficulty here would be initiating the decryption before the initramfs is available.

Authenticity of the server and integrity as well as confidentiality of the data here rely solely on the fact that the encryption key is only known to the server.

\subsubsection{Initrd in Unified Kernel Image}
Alternatively, the initramfs could be integrated into a UKI (see \Vref{subsec:systemd-stub}) and then signed together with it.
This is the easiest way, however, changing the initramfs would also require recreating and signing the UKI again.

Here, the same guarantees as for the kernel would be given, the server's authenticity and integrity of the data.

\subsubsection{TPM key sealing}
Last but not least, no further protection for the initramfs could be used.
Instead, the integrity of the key within the TPM could be guaranteed by binding it to certain PCR values, also called key sealing.
Two requirements need to be given to provide this guarantee:
the key can only be retrieved from the TPM if the right initramfs was loaded and
the rest of the process can only continue with the key.
A problem could arise when a change of the initramfs or even the kernel render the key in the TPM unusable and require another provisioning.

If all requirements are met, the server's authenticity and the integrity of the data can be considered guaranteed.

\subsection{iPXE scripts}\label{subsec:ipxe-scripts}
Since Secure Boot cannot protect the iPXE script, this section discusses the risk of an unprotected script and shows an option for securing it.
By changing the iPXE script, an attacker is able to modify the executed kernel, the used initramfs and the parameters the kernel is run with.
The attacker will be prevented from modifying kernel or initramfs if Secure Boot is used with one of the options described above to protect the initramfs, since invalid files would just be rejected.
Hence, only the kernel parameters are an actual attack vector that could be used for various attacks\footurl{https://sysdig.com/blog/kernel-parameters-falco/}.

The necessary protection depends on how the initramfs is protected:
\begin{itemize}
    \item If the initramfs is \textbf{integrated in a UKI}, the kernel parameter could also be integrated and therefore protected by Secure Boot.
    In that case, additional kernel parameters provided outside the integrated image will be ignored.\footurl{https://manpages.debian.org/bookworm/systemd/systemd-stub.7.en.html}
    However, it is not always feasible to integrate the kernel parameters with the unified kernel image,
    because currently the IP configuration for the host system is included in the kernel parameters
    to save the time of a repeated DHCP request.
    Additionally, it is planned to implement a feature to allow better device access within the virtual machines
    for the computer pool use case, which required device-specific kernel parameters to be set.
    In that case, it would be necessary to create a unified image for every possible device configuration.
    \item If \textbf{TPM key sealing} is used to guarantee the integrity, there is no need for further measures
    as changes in the kernel parameters also change the PCR state and therefore would render the key in the TPM unusable.
    \item If the initramfs is protected by \textbf{disk encryption using a TPM key}, further security measures
    for the kernel parameters are necessary, since this encryption cannot be used for the parameters.
\end{itemize}

\subsubsection*{iPXE signing}
Therefore, the iPXE script only needs to be secured additionally if the initramfs is encrypted using a TPM key or
if the kernel parameters are too variable to integrate them with kernel and initramfs.

In these cases, the integrity could be secured using iPXE signing\footurl{https://ipxe.org/crypto}.
This process is similar to Secure Boot as it also secures the integrity of binaries (and in this case also iPXE scripts)
by signing them with a key whose public part is known to the client.
Other than Secure Boot, where the key is stored in the UEFI, here the key needs to be embedded in the iPXE binary.
Also, iPXE signing protects only binaries and scripts that are loaded within the iPXE\@.
Additionally, in the binary, a flag is set that only signed binaries and scripts are to be accepted.
Then, every script and binary can be signed by the server and verified by the client.

\section{Within the host system}\label{sec:within-host-system}
Now that the host system has started,
the next step is to secure the transmission of information over dnbd3 in steps 6 and 7.
This task can be divided in two problems:
first of all, a key needs to be distributed between server and client without exposing it to a possible attacker.
Secondly, this key then has to be used for some encryption of the data.
For the key distribution, TPM can be used as explained in the following subsection.

\subsection{TPM}\label{subsec:tpm}
As explained in \Vref{subsec:trusted-platform-modules},
TPM is a hardware module that can store secrets (Root of Trust for Storage)
and report about the software's integrity (Root of Trust of Reporting).

One of its main purposes is to store encryption keys.
To use such key to secure the transmission,
the trust between server and TPM has to be established in a provisioning step.
It could be done by creating an asymmetric key pair in the TPM and transmitting the public key to the server.

After this is achieved, unauthorized access to the private key in the TPM should be prevented.
Three possible approaches to that will be presented in the following paragraphs.

\subsubsection{Key sealing}
The first option is to seal the keys using PCR constraints.
In that case, the key can only be used if the binaries used to boot the system
did exactly the same changes to the PCR as they did when the TPM was set up.
This very effectively prevents the key to be accessed when the operating system was tampered with.
However, it might also prevent access after, for example, a kernel update
and would then require a complete new provisioning.

\subsubsection{Password protection}
The second option could be using a passwort to restrain access to the key.
The main problem for this option would be to provide the password to the TPM
without enabling an attacker to receive this password as well.
This would therefore create another key distribution problem.

\subsubsection{Trust in software integrity}
Lastly, if the system has enabled secure boot, there is already protection of the software's integrity in place.
Hence, one might trust in the fact that no unauthorized access to the TPM can happen,
since no unauthorized software is allowed to boot on the system.
This option only offers lower security than the previous options when Secure Boot can be circumvented —
for example because of a leaked signing key —, but no root access using the untampered boot images is possible.

\subsection{Encryption}\label{subsec:encryption}
As already described in \Vref{subsec:computer-labs,subsec:data-centers},
encryption is so far only used in the data center environment to decrypt a config file.
The corresponding key is stored in a tag of the Data Center Manageability Interface (DCMI).
This interface is only accessible with root privileges, though it is not designed for secure data storage.

To improve the security of the key storage in the data center use case or
introduce a secure key storage in the computer lab use case, the secure enclave of a
Trusted Platform Module (TPM)~\cite{tcg2019} could be used if the client has such a module.

For this approach, at the initial configuration a public/private key pair is generated in the TPM\@.
And the public key is sent to the central server, encrypted with the servers public key, for example.
This could be implemented by a special deployment image or script which is run at the initial configuration.

At the further boots, the server encrypts a per-session symmetric encryption key using the clients public key
and sends it to the client where the TPM can be used to decrypt the key.
In this case, the identity of the client is verified by sending something that can only be encrypted with its private key.
The identity of the server, on the other hand, is only verified by the knowledge of the client's public key.
This is not a very strong protection as the public key should be considered public — although it is not in this case.

Finally, the key that is transferred as described before can be used for en- and decrypting the data
sent over the network, mainly over the dnbd3 protocol, but also other encrypted data is possible.
This could be done, for example, using the cryptography module of the device mapper, dm-crypt.
Additionally, this key could then also be used to decrypt an encrypted configuration file
as it is already used in the data center use case
or a SquashFS as could be observed in the computer lab use case with the minilinux host.
In the context of this work, the details of encrypting the transferred data will not be discussed further.

    \chapter{Implementation and Evaluation}\label{ch:experiments}
After multiple different approaches were described in the preceding chapter, this chapter gives an explanation of why
certain approaches were chosen for the implementation in the first section and describe said implementation in the second section.
A collection of all code used here can also be found in an online Git repository\footurl{https://naclador.de/mosers/stateless-boot-secure-code}.
In the last section, measurements of the boot times for different configurations are visualized and discussed.

\section{Chosen approaches}\label{sec:chosen-approaches}
\subsection*{Boot to iPXE}
For this step, Secure Boot (see \Vref{subsec:pxe-secure-boot}) was chosen for the test implementation,
since it offers all necessary security guarantees while only requiring little manual work (for enrolling the custom certificates).
Additionally, it can be used on top of any boot method like PXE boot, UEFI HTTP boot or USB boot, so it gives higher flexibility.

\subsection*{Host system boot from iPXE}
Since Secure Boot was chosen for the last step, it will also be used for this step.
The integrity of both the initramfs and the kernel parameters will be guaranteed by integrating it in the boot binary and signing it altogether.
As in some cases, it is not effective to integrate the kernel parameters into a unified kernel image,
the implementation of iPXE signing is also described.

\subsection*{Within host system}
Within host system, the TPM will be used for key storage.
The satellite server will then use the public part of this key to securely transmit a symmetric key to the client.
Subsequently, the symmetric key can be used for de- and encrypting the transmitted data.

\section{Proof-of-concept implementation}\label{sec:poc-implementation}
After describing the setup of a lab network for testing the implementation,
the following subsections contain each a step of the implementation,
together with a paragraph describing the necessary efforts of implementing it,
compared to the requirements listed in \Vref{sec:pb-requirements}.
In this context, the term server will refer to
the satellite server (for the computer lab use case) and the boot selection service (for the data center use case).

\subsection{Lab network setup}\label{subsec:poc-lab}

To create a proof-of-context implementation, a small lab network was set up.
It contained the following components:
\begin{itemize}
    \item a router (10.0.2.1/24)
    \item PC 1 used as virtual machine host (10.0.2.11) with multiple virtual machines connected to the router with interface bridging:
    \begin{itemize}
              \item DHCP Server based on Ubuntu Server 20.04.2 LTS and the ISC DHCP server 4.4.1 (10.0.2.2)
			  \footnote{As of the end of 2022, ISC has announced the end of maintenance for the ISC DHCP server
              which is replaced by ISC Kea (see \url{https://www.isc.org/dhcp/}).
              For the proof-of-concept, however, the resulting security concerns can be disregarded.}
              \item Satellite Server based on bwLehrpool server 3.10 (10.0.2.3)
    \end{itemize}
    \item PC 2 used as PXE Client (10.0.2.42)
\end{itemize}

When EFI boot was enabled in VirtualBox 7.0, PXE boot was not possible anymore.
With the \texttt{virtio-net} network adapter type selected, PXE boot started but threw an error,
while the other adapter types did not start PXE boot at all.
Enabling TPM and Secure Boot changed the type of error that was show.
This is expected to be because EFI mode is only experimental in VirtualBox\footurl{https://docs.oracle.com/en/virtualization/virtualbox/7.0/user/BasicConcepts.html\#efi}.
Therefore, it was necessary to create a lab network that could contain both virtual and physical machines
and then use a physical machine as the client.

\subsection{Secure Boot}\label{subsec:poc-secure-boot}
To verify the binaries used in the boot process, it is necessary to first create multiple key pairs
and then use them to sign the used binaries~\cite[Creating Secure Boot Keys]{smith2015}.
Both can be done on the server since every server should have its own signing key.
The used commands are shown in the following two Listings.

\Vref{lst:sb-mkkeys} shows the script \texttt{mkkeys.sh}.\footurl{https://www.rodsbooks.com/efi-bootloaders/mkkeys.sh}
It requests the input of a common name and then creates all certificates signs the certificates with each other
as listed in \Vref{subsec:secure-boot}.
Additionally, it creates all files necessary to enroll the certificates into the UEFI\@.

\begin{tcbcode}[label=lst:sb-mkkeys]{Secure Boot - \texttt{mkkeys.sh}}
#!/bin/bash
# Copyright (c) 2015 by Roderick W. Smith
# Licensed under the terms of the GPL v3

echo -n "Enter a Common Name to embed in the keys: "
read NAME

openssl req -new -x509 -newkey rsa:2048 -subj "/CN=$NAME PK/" \
        -keyout PK.key -out PK.crt -days 3650 -nodes -sha256
openssl req -new -x509 -newkey rsa:2048 -subj "/CN=$NAME KEK/" \
        -keyout KEK.key -out KEK.crt -days 3650 -nodes -sha256
openssl req -new -x509 -newkey rsa:2048 -subj "/CN=$NAME DB/" \
        -keyout DB.key -out DB.crt -days 3650 -nodes -sha256
openssl x509 -in PK.crt -out PK.cer -outform DER
openssl x509 -in KEK.crt -out KEK.cer -outform DER
openssl x509 -in DB.crt -out DB.cer -outform DER

GUID=`python3 -c 'import uuid; print(str(uuid.uuid1()))'`
echo $GUID > myGUID.txt

cert-to-efi-sig-list -g $GUID PK.crt PK.esl
cert-to-efi-sig-list -g $GUID KEK.crt KEK.esl
cert-to-efi-sig-list -g $GUID DB.crt DB.esl
rm -f noPK.esl && touch noPK.esl

sign-efi-sig-list -t "$(date --date='1 second' +'
                  -k PK.key -c PK.crt PK PK.esl PK.auth
sign-efi-sig-list -t "$(date --date='1 second' +'
                  -k PK.key -c PK.crt PK noPK.esl noPK.auth
sign-efi-sig-list -t "$(date --date='1 second' +'
                  -k PK.key -c PK.crt KEK KEK.esl KEK.auth
sign-efi-sig-list -t "$(date --date='1 second' +'
                  -k KEK.key -c KEK.crt db DB.esl DB.auth

chmod 0600 *.key

echo ""
echo "For use with KeyTool, copy the *.auth and *.esl files to"
echo "a FAT USB flash drive or to your EFI System Partition (ESP)."
echo "For use with most UEFIs' built-in key managers, copy the"
echo "*.cer files; but some UEFIs require the *.auth files."
echo ""
\end{tcbcode}

The code in \Vref{lst:sb-setup} installs the necessary tool and used the script \texttt{mkkeys.sh} described before to create all keys.
Afterwards, the command \texttt{sbsign}\footurl{https://manpages.debian.org/bookworm/sbsigntool/sbsign.1.en.html}
is used to sign the iPXE binary and the kernel.
The \texttt{sbsign} command expects the private key from the signature database in ASCII-armored format at \texttt{--key},
the matching PEM certificate at \texttt{--cert} and the binary to sign as a positional argument.
It then creates the signature, appends it to the binary and saves the signed binary to the location given at \texttt{--output}.

\begin{tcbcode}[label=lst:sb-setup]{Secure Boot - creating keys and signing binaries}
apt install efitools
mkdir ~/sbkeys && cd ~/sbkeys
wget https://www.rodsbooks.com/efi-bootloaders/mkkeys.sh
chmod +x mkkeys.sh
./mkkeys.sh
cd /srv/openslx/tftp
mv ipxe.efi ipxe.efi.bak
sbsign --key ~/sbkeys/DB.key --cert ~/sbkeys/DB.crt ipxe.efi.bak --output ipxe.efi
cd /srv/openslx/www/boot/bwlp/maxilinux-u2004/31r1
mv kernel kernel.bak
sbsign --key ~/sbkeys/DB.key --cert ~/sbkeys/DB.crt kernel.bak --output kernel
\end{tcbcode}

Subsequently, it is necessary to copy the created \texttt{.cer} files from \texttt{sbkeys} onto a USB stick and enroll them into the UEFI loader.
For the proof-of-concept, this was done manually using the UEFI menu.

\subsubsection*{Using LockDown.efi}
However, enrolling can also be done with a special boot image~\cite[Securing Multiple Computers]{smith2015}
which is recommended for the use with a higher number of clients.
In that case, lines 1 to 5 of the preceding code listing need to be replaced with the following code,
which installs tools necessary to build the \texttt{efitools} program, clones it from the kernel's git repository
and then builds it, which also creates the keys that were created with the script \texttt{mkkeys.sh} before.

\begin{tcbcode}{Secure Boot - build efitool with custom keys and enrollment binary}
apt install gnu-efi help2man sbsigntool
cpan File::Slurp
git clone git://git.kernel.org/pub/scm/linux/kernel/git/jejb/efitools
cd efitools && make
\end{tcbcode}

Now all keys can be found in the \texttt{efitools} directory, so the signing process needs to be changed accordingly.
Then the \texttt{LockDown.efi} binary can be used to make a bootable USB stick, for example, that enrolls those keys.
For testing purposes, multiple binaries called \texttt{*-signed.efi} can be found in the directory.
These binaries should work even after Secure Boot is enabled on the client.

\subsubsection*{Necessary efforts}
For both use cases, this step has to be done only once, so it can be integrated in the set-up of the server.

Manual work is only required for enrolling the keys into UEFI, for which a custom image has to be booted once.
In the data center use case, it should be possible to change the boot medium once automatically.
However, if it is not possible to boot the custom image for enrolling the keys over the network in the computer lab use case, every machine might have to be changed manually.
In this case, it is also possible to use clients with Secure Boot enabled and disabled with the same satellite server,
as clients with either setting accept signed images, so implementation could be done gradually.
\newpage

\subsection{Unified Kernel Image (UKI)}\label{subsec:poc-uki}
The binaries signed for Secure Boot in the previous subsection only include the kernel aside the iPXE binary.
Since it is required to include the initramfs and the kernel parameters into a UKI to be able to sign them as well,
this will be done in the following subsection.
As multiple initramfs cannot be easily unified into one image,
the second initramfs \texttt{bwlp.cpio} was skipped for this example as it only contains the splash image.

At first, two files are created containing the kernel parameters that are sent in the iPXE script until now
and the OS release information for the signed binary which can be extracted from the host system:

\begin{tcbcode}{Unified Kernel Image - \texttt{/srv/cmd}}
slxbase=boot/bwlp/maxilinux-u2004/31r1 slxsrv=10.0.2.3 slx.stage4.path=stage4/bwlp/maxilinux-bookworm-6.1.33-94.qcow2 bridged quiet nosplash systemd.show_status=0 rd.shell=0 rd.emergency=reboot ipv4.ip=10.0.2.42 ipv4.router=10.0.2.1 ipv4.dns=10.0.2.1 ipv4.hostname=client ipv4.if=a8:a1:59:0b:fe:87 ipv4.ntpsrv=de.pool.ntp.org ipv4.subnet=255.255.255.0 slx.swap ibt=off slx.ipxe.id=1
\end{tcbcode}

The kernel parameters contain network details that differ for every client,
so it should be checked whether they can be transferred to the client differently.
The other kernel parameters are set per boot image and hence can be integrated into the binary.

\begin{tcbcode}{Unified Kernel Image - \texttt{/srv/osrel}}
PRETTY_NAME="Debian GNU/Linux 12 (bookworm)"
NAME="Debian GNU/Linux"
VERSION_ID="12"
VERSION="12 (bookworm)"
VERSION_CODENAME=bookworm
ID=debian
HOME_URL="https://www.debian.org/"
SUPPORT_URL="https://www.debian.org/support"
BUG_REPORT_URL="https://bugs.debian.org/"
\end{tcbcode}

After the prerequisites are done, the image can be unified and signed from within the directory
containing the kernel and initramfs.
The unifying is done using the \texttt{objcopy}\footurl{https://man.freebsd.org/cgi/man.cgi?query=objcopy} utility,
which is able to copy the content of multiple object files into one.
The argument \texttt{--add-section} defines different sections of the new object file,
while \texttt{--change-section-vma} manually sets the memory addresses for each section.
Additionally, to the previously described OS release information and kernel parameters,
the unsigned kernel is added as well as the initramfs.
These parts are all added to the systemd-stub which was previously explained in \Vref{subsec:systemd-stub}.

\begin{tcbcode}{Unified Kernel Image - create and sign}
cd /srv/openslx/www/boot/default
objcopy \
--add-section .osrel=/srv/osrel \
--change-section-vma .osrel=0x20000 \
--add-section .cmdline=/srv/cmd \
--change-section-vma .cmdline=0x30000 \
--add-section .linux=kernel.bak \
--change-section-vma .linux=0x40000 \
--add-section .initrd=initramfs-stage31 \
--change-section-vma .initrd=0x3000000 \
/usr/lib/systemd/boot/efi/linuxx64.efi.stub kernel.efi.bak
sbsign --key ~/sbkeys/DB.key --cert ~/sbkeys/DB.crt kernel.efi.bak --output kernel.efi
\end{tcbcode}

Finally, the iPXE script delivered to the client has to be changed to refer to the newly created \texttt{kernel.efi}.
For the purpose of the proof-of-concept, this was done by creating an iPXE script file and modifying the PHP script,
so it delivers that iPXE script, as shown in the following Listings.
The content of the iPXE script is the same as the script delivered by the web application,
just the \texttt{boot} command is changed to boot only a single image without any additional initramfs or kernel parameters.
In a production implementation, it needs to be professionally integrated into the satellite web server application.

\begin{tcbcode}{Unified Kernel Image - \texttt{/srv/script.ipxe}}
#!ipxe
set ipappend1 ip=${ip}:10.0.2.3:${gateway}:\${netmask}
set ipappend2 BOOTIF=01-${mac:hexhyp}
set serverip 10.0.2.3 ||
iseq ${idx} ${} && set idx:string X ||
iseq ${self} ${} && set self http://10.0.2.3/boot/ipxe? ||
set menuentryid 1 ||
imgfree ||
boot /boot/default/kernel.efi || goto fail
goto fail
goto end
:fail
prompt --timeout 5000 Error launching selected boot entry ||
:end
\end{tcbcode}

After the updated iPXE script is created, it is necessary to modify the web server
at the file \texttt{/srv/openslx/www/slx-admin/modules/serversetup/api.inc.php} to deliver the script:

\begin{tcbcode}[listing options={firstnumber=18, language=diff}]{Unified Kernel Image - modify web server to deliver custom script}
    } elseif ($entryId !== false) {
        $entry = MenuEntry::get($entryId);
        $data = $builder->getMenuEntry($entry);
+       if ($entryId === 1) {
+           $data = file_get_contents("/srv/script.ipxe");
+       }
    } else {
\end{tcbcode}

After completing all steps, the PXE client was able to boot into the created kernel image using Secure Boot.

\subsubsection*{Necessary efforts}
Creating and signing a unified image is necessary for every change in the kernel, the initramfs or the kernel parameters.
In both use cases, creating the kernel or initramfs is automated, so this step can be added there without manual effort.

The changes in the iPXE script showed here are only for proof-of-concept purposes.
For a productive implementation, the generation of iPXE scripts including the necessary changes should be
integrated better in the SLX server instead of hard-wiring a static script to a specific boot option.
However, a high effort for development is not expected.
Afterwards, no changes to the build and boot process are necessary.

\subsection{iPXE signing}\label{subsec:ipxe-signing}
If the kernel parameters or the initramfs cannot be integrated with the kernel,
their integrity can also be protected by using iPXE signing as described in the iPXE documentation\footurl{https://ipxe.org/crypto}.

As shown in \Vref{lst:ipxe-sign-setup}, initially a key, a certificate and a basic configuration file
for a certificate authority (CA) are created.
It is important to set the \texttt{keyUsage} fields so the resulting keys can be used for the intended signing purposes.
Afterwards, \texttt{openssl} is used to create the keys,
while \texttt{-extensions codesigning} is set to enable the intended purpose again.
Lastly, the iPXE binary is built with the newly created CA certificate included and set to trusted.

\begin{tcbcode}[label={lst:ipxe-sign-setup}]{iPXE signing - create a signing key and build binary with it}
openssl req -x509 -newkey rsa:2048 -out ca.crt -keyout ca.key -days 1000
echo 01 > ca.srl
touch ca.idx
mkdir signed
cat << EOF >> ca.cnf
[ ca ]
default_ca             = ca_default

[ ca_default ]
certificate            = ca.crt
private_key            = ca.key
serial                 = ca.srl
database               = ca.idx
new_certs_dir          = signed
default_md             = default
policy                 = policy_ipxe
preserve               = yes
default_days           = 90
unique_subject         = no

[ policy_ipxe ]
commonName             = ipxe.ca
countryName             = match
stateOrProvinceName     = match
organizationName        = match
organizationalUnitName  = optional
commonName              = optional
emailAddress            = optional

[ cross ]
basicConstraints       = critical,CA:true
keyUsage               = critical,cRLSign,keyCertSign

[ codesigning ]
keyUsage                = digitalSignature
extendedKeyUsage        = codeSigning
EOF

openssl req -newkey rsa -keyout codesign.key -out codesign.req
openssl ca -config ca.cnf -extensions codesigning -in codesign.req -out codesign.crt
make bin/undionly.kkkpxe CERT=ca.crt TRUST=ca.crt
\end{tcbcode}

Additionally, each iPXE script (the embedded script and scripts generated by SLX) needs to be changed.
At the beginning of the embedded iPXE script,
the command \texttt{imgtrust --permanent} should be added to enforce using verified files.
If \texttt{imgtrust} is set, no files are accepted before being verified with \texttt{imgverify}.
Every occurrence of the command \texttt{chain} (aliases are \texttt{imgexec} and \texttt{imgload}) should be changed,
so that the signature of the file is requested and verified before execution.
The \texttt{imgfetch} command downloads a file into the memory and the \texttt{--name} parameter creates a simplified name for this file.
It is then verified by the \texttt{imgverify} command, which expects the name of the file to verify
and then the location of the separate signature file.
Finally, the file can be loaded using the \texttt{imgload} command.

\begin{tcbcode}[listing options={language=diff}]{iPXE signing - example of changing a script}
    set self:string http://10.0.2.3/boot/ipxe?uuid=${uuid}&mac=${mac}&manuf=${manufacturer:uristring}&product=${product:uristring}&platform=${platform:uristring}&slx-extensions=${slxext}
-   chain -ar ${self} || goto fail
+   imgfetch --name ipxe ${self}
+   imgverify ipxe ${self}&sig=true
+   imgload ipxe || goto fail
\end{tcbcode}

To make this work, every static file (e.g.,\ binaries) should have a signature file stored in the same directory.
For files that are dynamically created (e.g.,\ scripts created by SLX),
the signature should be created when a script is requested with certain parameters and then be sent to the client
when the same parameters with an additional signature parameter are requested.

The command for creating a signature on the command line is as show in \Vref{lst:ipxe-sign}.
The \texttt{openssl cms}\footurl{https://www.openssl.org/docs/man1.0.2/man1/cms.html} command
takes an unsigned binary and the necessary keys and creates a separate S/MIME signature file.

\begin{tcbcode}[label={lst:ipxe-sign}]{iPXE signing - sign a file}
openssl cms -sign -binary -noattr -in kernel.efi \
    -signer codesign.crt -inkey codesign.key -certfile ca.crt \
    -outform DER -out kernel.efi.sig
\end{tcbcode}

Since the signature is not appended to the file, there is no collision with Secure Boot signatures.
In this case, the iPXE signature needs to be created after the Secure Boot signature,
as the Secure Boot signature changes the file hash again and would render an already created iPXE signature invalid.

\subsubsection*{Necessary efforts}
This method requires the generation of certificates when the server is set up and the iPXE binary is build.
This can be added to the current set-up process and does not require additional manual work.

Furthermore, the generation of iPXE scripts needs to be changed in a way that on the one hand,
the process of verifying files is added to the script, and on the other hand,
the signature of every script can be retrieved by the iPXE client.
It requires a more complex modification of the SLX server with a higher implementation effort.

\subsection{TPM}\label{subsec:poc-tpm}
To use the TPM for key distribution, the most important part is a robust provisioning~\cite{segall2012provisioning}.
Any threat at first use should be prevented,
therefore, the system should be booted with a trusted medium like USB or CD\@.
As long as the TPM is not used as Root of Trust for Reporting, no Endorsement Key (EK) is required.
Additionally, setting a secret owner authorization should not be necessary
since no additional security is expected from that in this use case.

Hence, mainly the creation of an asymmetric key pair and the transmission of the public key to the server is critical.
It is strongly recommended to use another trusted medium for the key transfer, like USB in the following example.
After installing the necessary tools and folders, the code in \Vref{lst:tpm-setup} uses the tool
\texttt{tpm2\_createprimary}\footurl{https://tpm2-tools.readthedocs.io/en/latest/man/tpm2_createprimary.1/}
to create a key pair stored in the TPM, which uses the asymmetric algorithm RSA2048 (\texttt{--key-algorithm=rsa2048}),
and saves the data into a file (\texttt{--key-context=key.ctx}),
that is necessary for the TPM to unambiguously identify the key (key context~\cite[section 4.17]{tcg2019}).
Consequently, \texttt{tpm2\_evictcontrol}\footurl{https://tpm2-tools.readthedocs.io/en/latest/man/tpm2_evictcontrol.1/}
is used to persistently store the key in the TPM, which makes it available under a given context handle number.
The handle range \texttt{81000000}$_{16}$–\texttt{810000FF}$_{16}$ is reserved for storage keys~\cite{tcg2019handles}.
Finally, \texttt{tpm2\_readpublic}\footurl{https://tpm2-tools.readthedocs.io/en/latest/man/tpm2_readpublic.1/}
is used to export this key's public key in the PEM format.

\begin{tcbcode}[label={lst:tpm-setup}]{TPM - prepare environment and create keys}
apt install tpm2-tools
mkdir usb && mount /dev/sdb1 usb && cd usb
tpm2_createprimary --key-algorithm=rsa2048 --key-context=key.ctx
tpm2_evictcontrol --object-context=key.ctx 0x81000001
tpm2_readpublic --object-context=0x81000001 --format=pem --output=key.pem
\end{tcbcode}

Only the public key \texttt{key.pem} needs to be saved for later use, together with the context handle, e.g.\ \texttt{0x81000001}.

After provisioning, the key in TPM can be used for en- and decryption, even after rebooting.
The commands \texttt{tpm2\_rsaencrypt}\footurl{https://tpm2-tools.readthedocs.io/en/latest/man/tpm2_rsaencrypt.1/}
and \texttt{tpm2\_rsadecrypt}\footurl{https://tpm2-tools.readthedocs.io/en/latest/man/tpm2_rsadecrypt.1/}
both take the key context handle as an argument, read data from STDIN or a file provided as positional argument
and write the result to STDOUT or an output file provided with \texttt{--output}.
Encryption of data for the TPM can also be done, for example, with the \texttt{openssl pkeyutl} command.
\footurl{https://www.openssl.org/docs/man1.1.1/man1/openssl-pkeyutl.html}

\begin{tcbcode}{TPM - en-/decrypt using RSA}
# Encryption using the TPM
echo "secret" | tpm2_rsaencrypt --object-context=0x81000001 --output=msg.enc
# Encryption using the public key
echo "secret" | openssl pkeyutl -encrypt -pubin -inkey key.pem -out msg2.enc
# Decryption using the TPM
tpm2_rsadecrypt --object-context=0x81000001 msg.enc
tpm2_rsadecrypt --object-context=0x81000001 msg2.enc
\end{tcbcode}

\subsubsection*{Necessary efforts}
First of all, for this implementation a TPM module needs to be available in the client
which cannot be taken as granted in the data center use case.

For the provisioning, it is necessary to run a prepared image on every client.
For the data center use case, it should be possible to boot a trusted medium remotely over the management interface.
In the computer lab use case, the most secure way would be booting every client manually over a trusted medium.
Alternatively, it would also be possible to start the provisioning from the satellite server.
The trust in the integrity of the provisioning could still be established by using Secure Boot.

Nevertheless, in both use cases, it is necessary to transmit the public key back to the server in a way
that both integrity and confidentiality of the clients public key can guaranteed.
The most secure way would be over a USB medium, but it also requires manual work on every client.
Alternatively, another secure channel like a secure provisioning network could be used.
Last but not least, the principle of \enquote{Trust on First Use} could be applied,
which means that the data is transmitted in unsecured clear text once but is still trusted afterwards.
Both the secure channel or unsecured clear text require no manual work on every client.

\section{Evaluation}\label{sec:evaluation}
\subsection{Boot time measurement}\label{subsec:boot-time-measurement}

For this section, the boot times for three different implementation levels and two different host system variants were measured.
Every measurement was done three times and the average was included here.
The UEFI's POST delay time of 10 seconds was subtracted from all measurements while the iPXE selection screen timeout
of 3 seconds was only subtracted from the measured boot time to the host system log-in screen.

The different implementation levels are:
\begin{itemize}
    \item \textbf{No Secure Boot:} The system is in the state directly after setup, no additional security technologies are implemented yet.
    \item \textbf{Basic Secure Boot:} Secure Boot is enabled, but only the necessary binaries (iPXE and kernel) are signed.
     The binaries do not contain any modifications apart from signatures.
    \item \textbf{Secure Boot with UKI:} Secure Boot is enabled, and the binaries loaded by iPXE are combined to a Unified Kernel Image (UKI).
\end{itemize}

The measurements were done for two different host system variants:
\begin{itemize}
    \item minilinux: \texttt{bwlp/minilinux-u2004/29r3}
    \item maxilinux: \texttt{bwlp/maxilinux-u2004/31r1}
\end{itemize}

\Vref{fig:boot-time-ipxe} shows the boot time until the iPXE selection screen is displayed for the combinations of the implementation levels and host system variants above.
\Vref{fig:boot-time} shows the boot time until the host system log-in screen is displayed for the same combinations.

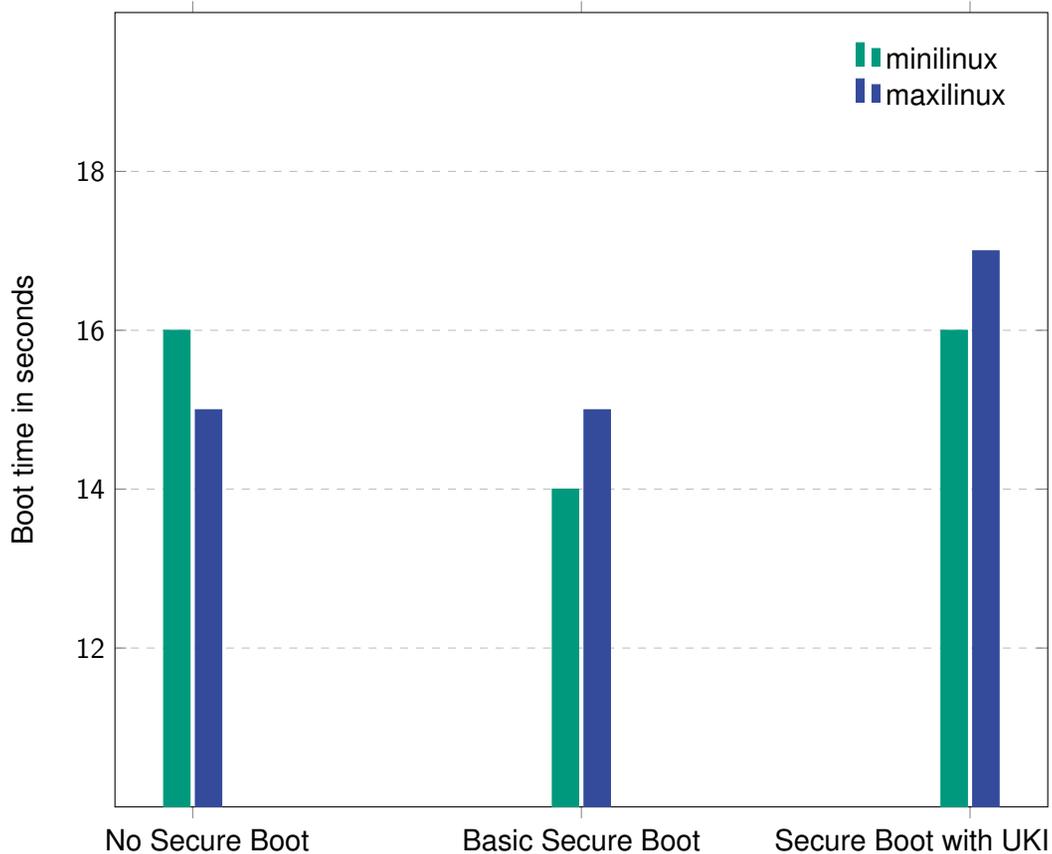
\begin{figure}[htb!]
\begin{center}
\begin{tikzpicture}
  \begin{axis}[
	ybar,
	width=1\textwidth,
    ylabel={Boot time in seconds},
    ymin=10, ymax=20,
    ytick={12,14,16,18},
    xtick={1,2,3},
    xticklabels={\quad{}No Secure Boot, Basic Secure Boot, Secure Boot with UKI\qquad\qquad\ },
    legend cell align=left,
	legend pos=north east,
	legend style={draw=none},
    ymajorgrids=true,
    grid style=dashed,
    legend entries={minilinux, maxilinux},
    tick label style = {font=\sansmath\sffamily},
    every axis label = {font=\sansmath\sffamily}
  ]
  \addplot[color=UniGruen,fill=UniGruen] table[x=Name, y=pxemini] {short.txt};
  \addplot[color=UniBlau,fill=UniBlau] table[x=Name, y=pxemaxi] {short.txt};
  \end{axis}
\end{tikzpicture}
\caption{Boot time to iPXE selection screen by implementation level and host system variant}
\label{fig:boot-time-ipxe}
\end{center}
\end{figure}
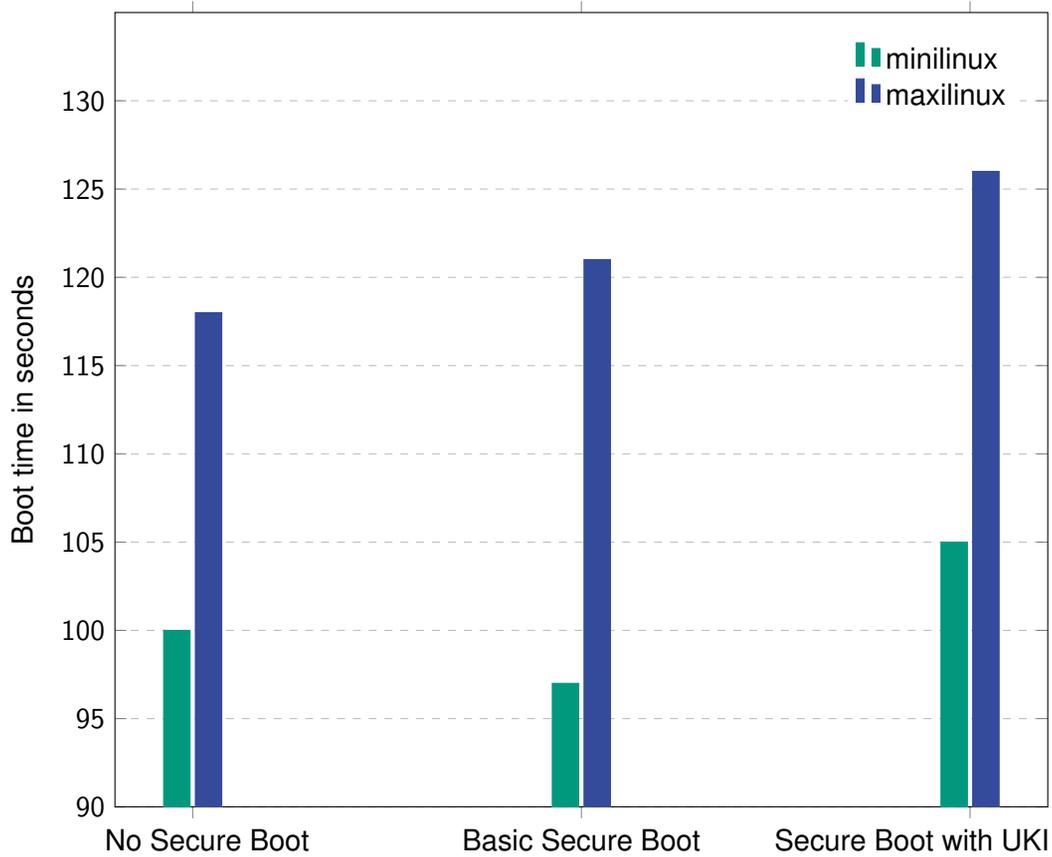
\begin{figure}[htb!]
\begin{center}
\begin{tikzpicture}
  \begin{axis}[
	ybar,
	width=1\textwidth,
    ylabel={Boot time in seconds},
    ymin=90, ymax=130,
    ytick={90,95,100,105,110,115,120,125,130},
    xtick={1,2,3},
    xticklabels={\quad{}No Secure Boot, Basic Secure Boot, Secure Boot with UKI\qquad\qquad\ },
    legend cell align=left,
	legend pos=north east,
	legend style={draw=none},
    ymajorgrids=true,
    grid style=dashed,
    legend entries={minilinux, maxilinux},
	enlarge y limits  = {upper,value=0.125},
    tick label style = {font=\sansmath\sffamily},
    every axis label = {font=\sansmath\sffamily}
  ]
  \addplot[color=UniGruen,fill=UniGruen] table[x=Name, y=minilinux] {short.txt};
  \addplot[color=UniBlau,fill=UniBlau] table[x=Name, y=maxilinux] {short.txt};
  \end{axis}
\end{tikzpicture}
\caption{Boot time to host system log-in screen by implementation level and host system variant}
\label{fig:boot-time}
\end{center}
\end{figure}

\subsection{Interpretation of the measurements}\label{subsec:interpretation-of-the-measurements}
As it can be seen in \Vref{fig:boot-time-ipxe}, the boot times for the iPXE binary vary by 3 seconds between 14 and 17 seconds.
No difference is to be expected between both the minilinux and maxilinux variants and
between the Basic Secure Boot and Secure Boot with UKI configuration,
as the boot process does not differ until the iPXE selection screen.
Therefore, it can be assumed that the differences are due to inaccuracies in the measurement,
varying network loading times and other reasons that are not related to the implemented mechanisms.

In \Vref{fig:boot-time} it can be observed that the boot times increase slightly with basic Secure Boot for maxilinux,
while they drop slightly for minilinux.
Since the binaries that have to be signed with basic Secure Boot
and checked by the UEFI are the same for minilinux and maxilinux,
it can be assumed that these are also pure deviations in the measurement that are not due to the implementation.
This means that these measurements do not show a significant increase in boot time
through the use of Secure Boot without UKI per se.

For the Secure Boot with UKI, the boot times increase by about 5 seconds for both minilinux and maxilinux.
Here, the difference made by the implementation of UKI cannot be denied.
It can be assumed that the time delay caused by additional steps necessary
to load the kernel from a UKI outweighs the time gained by saved HTTP requests.
However, it could be observed that the boot process of minilinux changed through the use of UKI,
so there may be other reasons for the delay — for example, the embedded kernel parameters.

Since iPXE signing was not included in the proof-of-concept implementation, no measurements could be done for it.
Although, only a slight increase of boot time is expected when the signature is loaded and verified for every binary and script.

The time overhead of encrypting dnbd3 disks was not analyzed;
however, tests by Cloudflare\footurl{https://blog.cloudflare.com/speeding-up-linux-disk-encryption/}
suggest that the use of dm\_crypt still enables read rates similar to those of HDDs.
In addition, transferring blocks over the network and decrypting them occupies different resources
(CPU load versus network load), so the two processes do not interfere with each other.
Thus, it is expected that the added delay will not forfeit the usability.
Using the TPM itself is not expected to add any delay to the boot times.

    \chapter{Conclusion}\label{ch:conclusion}
Using Secure Boot and TPM, the stateless system remote boot can be enhanced to fulfill the security goals of
operating system integrity on the client and confidentiality of the transmitted secrets.
While the Secure Boot is used to only allow signed binaries to boot the system,
the TPM stores a private key that can then be used to decrypt transmitted data,
be it configuration tarballs or dnbd3 blocks.

The measurements done for Secure Boot show that Secure Boot itself adds no significant delay to the boot process.
A small delay was measured for the use of Unified Kernel Images.
A measurement for iPXE signing as well as for the accessing
an encrypted image over dnbd3 should be content of further works.

The effort to enable Secure Boot and integrate signing in the process
of building or loading binaries is only necessary once.
After that, manual intervention is no longer necessary.
Using UKI or iPXE signing also requires only a one-time implementation effort,
with UKI mainly increasing the time needed to (automatically) create binaries
after changes to kernel, initramfs or the kernel parameters,
while iPXE signing has only a higher effort to implement signing of scripts that are dynamically generated by SLX\@.

It should be further analyzed for both use cases whether it is feasible
to integrate initramfs and kernel parameters in the kernel.
If they should be integrated, the process of integrating and signing needs to be repeated for each change.
Additionally, client-specific modifications cannot be done anymore in initramfs or kernel parameters.
Here, it is suggested to check whether the information carried by those parameters can also be transmitted in a different way.
If the integration is not possible, it will be necessary to implement iPXE signing as described in \Vref{subsec:ipxe-signing}.

For the TPM, trusting in the integrity of the boot process only prevents some attacks.
An attacker with skill and motivation to tamper with the hardware
will still be able to use the stored key to decrypt data.

For the computer lab use case, the gain in security by storing the key in the TPM is clearly recognizable.
Since using the DCMI tag in the data center use case already provides some security, here the gain is not that clear.
Both ways of storing the key provide protection against an attacker without root privileges on the host,
insofar as the software used in each case does not contain any vulnerabilities.
If root privileges are gained on the host system, an attacker can use the key from both storages.
When a TPM is used, an attacker cannot retrieve the key, but only use the TPM to decrypt the symmetric key,
which means he can access this key while he has still an access to the system.
As soon as he does not have access anymore and the symmetric key is changed, he cannot decrypt data anymore.
Additionally, the DCMI only allows reading 16 Bytes at once,
which offers a key length of only 128 bits~\cite[Table 6--8, Get Asset Tag Command]{intelcorporation2011}.
This weakness can be fixed by requesting different offsets of the asset tag which has a maximum length of 63 Bytes,
which would allow a more secure key length of up to 504 bits.
Nevertheless, using the asset tag can only be a temporary solution since the manufacturer
could start using it for its intended purpose, which is to store serial numbers for support cases.
Therefore, the current security gain of using the TPM is low
and might not justify the high expenses of buying server hardware with a TPM, however,
continuing to use the DCMI asset tag as a key storage bears the risk that this storage might not be usable in the future.

The final step of encryption was not completely described in this thesis.
It has to be checked which ways of encryption work best in which situations.
Encrypting full disks transparently below the dnbd3 layer seems to be most useful in a general case,
since it is not necessary to examine which data needs confidentiality — for security or privacy reasons.
However, if data changes frequently and maybe is adapted to the client,
it would be more feasible to continue the practice of encrypting single files and transmitting them over HTTP\@.
    
    \bibliographystyle{ieeetr}
    \bibliography{bib/thesis_main}
    \addcontentsline{toc}{chapter}{Bibliography}
    \newpage
    \thispagestyle{empty}
    \mbox{}
    \newpage
    \thispagestyle{empty}
    \mbox{}
\end{document}